\documentclass[12pt,preprint]{aastex}
\usepackage{natbib}
\usepackage{amssymb}
\usepackage{amsmath}
\slugcomment{for \emph{Astrophysical Journal}, \today}

\shorttitle{Quenching on Cool Brown Dwarfs and Hot Jupiters}

\shortauthors{Visscher \& Moses}

\begin{document}

\title{Quenching of Carbon Monoxide and Methane in the Atmospheres of Cool Brown Dwarfs and Hot Jupiters}

\author{Channon Visscher$^{1}$, and Julianne I.~Moses$^{2}$}

\affil{$^{1}$Lunar and Planetary Institute, USRA, 3600 Bay Area Blvd, Houston, TX 77058}

\affil{$^{2}$Space Science Institute, 4750 Walnut St, Suite 205,
Boulder, CO 80301}

 \email{visscher@lpi.usra.edu, jmoses@spacescience.org}

\begin{abstract}
We explore $\textrm{CO}\rightleftarrows \textrm{CH}_{4}$ quench kinetics in the atmospheres of 
substellar objects using updated time-scale arguments, as suggested by a thermochemical kinetics 
and diffusion model that transitions from the thermochemical-equilibrium regime in the deep 
atmosphere to a quench-chemical regime at higher altitudes.  More specifically, we examine CO 
quench chemistry on the T dwarf Gliese 229B and CH$_{4}$ quench chemistry on the hot-Jupiter 
HD 189733b.  
We describe a method for correctly calculating reverse rate coefficients for chemical reactions, 
discuss the predominant pathways for $\textrm{CO}\rightleftarrows \textrm{CH}_{4}$ interconversion
as indicated by the model, and demonstrate that a simple time-scale approach can be used to accurately 
describe the behavior of quenched species when updated reaction kinetics and mixing-length-scale 
assumptions are used.  Proper treatment of quench kinetics has important implications for estimates 
of molecular abundances and/or vertical mixing rates in the atmospheres of substellar objects.  Our model results indicate significantly higher $K_{zz}$ values than previously estimated near the CO quench level on Gliese 229B, whereas current model-data comparisons using CH$_{4}$ permit a wide range of $K_{zz}$ values on HD 189733b.  We also use updated reaction kinetics to revise previous estimates of the Jovian water abundance, based upon the observed abundance and chemical behavior of carbon monoxide.  The CO chemical/observational constraint, along with \emph{Galileo} entry probe data, suggests a water abundance of approximately $0.51-2.6\times$ solar (for a solar value of H$_{2}$O/H$_{2}=9.61\times10^{-4}$) in Jupiter's troposphere, assuming vertical mixing from the deep atmosphere is the only source of tropospheric CO.  \end{abstract}


\section{Introduction}
In general, thermochemical equilibrium governs the composition of the deep atmospheres of 
giant planets and brown dwarfs because they are warm enough for chemical reactions to 
readily overcome energy barriers to reaction kinetics.  However, disequilibrium 
processes in substellar atmospheres are well known.  In addition to photochemistry driven by 
ultraviolet irradiation,
atmospheric mixing is one of the dominant mechanisms that drives the chemical composition out of equilibrium.  In this scenario, rapid vertical mixing may transport a parcel 
of gas to higher, cooler altitudes before its chemical constituents have had sufficient time to 
attain equilibrium via reaction chemistry --- a phenomenon that has been proposed to explain 
the overabundance of various ``disequilibrium'' species in the atmospheres of Jupiter, Saturn, 
Uranus, and Neptune 
\citep[e.g.,][]{prinn1976,prinn1977,barshay1978,fegley1979,prinn1981jgr,prinn1984,lewis1984,fegley1985apj,fegley1986apj,fegley1988,fegley1991,fegley1994,lodders1994,lodders2002,bezard2002,taylor2004,visscher2005,fouchet2009,visscher2010icarus,moses2010}, 
brown dwarfs \citep[e.g.,][]{fegley1996,noll1997,griffith1999,griffith2000,saumon2000,lodders2002,golimowski2004,saumon2006,saumon2007,visscher2006,leggett2007,hubeny2007,geballe2009,king2010,yamamura2010}, 
and extrasolar giant planets
\citep[e.g.,][]{cooper2006,fortney2006hd149,burrows2008,line2010,madhusudhan2011,stevenson2010,moses2011}.  

\citet{prinn1977} first developed an analytical model to explain the observed overabundance 
of CO in Jupiter's troposphere due to strong vertical mixing.  In this approach, a 
time scale for convective mixing ($\tau_{mix}$), based upon an estimated mixing length scale and 
vertical mixing rate, is compared to 
a time scale for chemical kinetics ($\tau_{chem}$), based upon an assumption about 
which chemical pathways will be important for interconversion between atmospheric constituents.   
At high temperatures in the deep atmosphere, thermochemical equilibrium is maintained because 
reaction kinetics operate faster than convective mixing ({\em i.e.,} $\tau_{chem}<\tau_{mix}$).    However, 
departures from equilibrium can occur at colder, higher altitudes when convective mixing begins 
to dominate over reaction kinetics ({\em i.e.}, when $\tau_{chem}>\tau_{mix}$), and the abundance of a molecular constituent may become  ``quenched'' at a value representative of the quench level 
(defined by $\tau_{chem}=\tau_{mix}$).  This level is different for each species \citep{fegley1985apj} 
and, in principle, \emph{every} species that is subject to reaction chemistry and atmospheric transport will quench if the appropriate time scale for reaction kinetics becomes longer than the time scale for convective mixing \citep[{e.g.}, see Fig.~6 in][]{visscher2010icarus}.  Here we investigate the chemical interconversion between CO and CH$_{4}$, which becomes quenched when $\tau_{chem}(\textrm{CO}\rightleftarrows\textrm{CH}_{4})>\tau_{mix}$ in a substellar atmosphere.  Whether either species is considered a ``disequilibrium'' species depends upon its quench abundance relative to its expected equilibrium abundance at higher altitudes.  For example, CO is present as a disequilibrium species on objects where CH$_{4}$ is the dominant carbon-bearing 
gas, including Jupiter and T dwarfs (such as Gliese 229B), whereas CH$_{4}$ is present as a disequilibrium species on objects where CO is the dominant carbon-bearing gas, including L 
dwarfs and hot Jupiters (such as HD 189733b and HD 209458b).

Although the general approach of \citet{prinn1977} is robust, uncertainties in identifying 
the rate-limiting step for $\textrm{CO}\rightleftarrows \textrm{CH}_{4}$ interconversion, errors in the 
treatment of reaction kinetics, and incorrect assumptions about the mixing length scale are 
found throughout the astrophysical literature for substellar objects.  As we discuss below, 
these issues may lead to incorrect conclusions regarding quench abundances for disequilibrium 
species and/or incorrect estimates of atmospheric mixing rates 
that are constrained by the disequilibrium chemistry.  In order to circumvent some of the uncertainties involved with traditional time-scale modeling approaches, we recently developed a 
one-dimensional photochemistry, thermochemistry, and transport model for the atmosphere of 
Jupiter \citep{visscher2010icarus,moses2010} and extrasolar giant planets \citep{moses2011}.  This approach allows us to accurately model the abundance of constituents in the transition 
region between the thermochemical regime in the deep troposphere  (where equilibrium is maintained 
via rapid reaction kinetics) and the quenched regime in the upper troposphere (where rapid 
atmospheric transport and slow reaction kinetics drive constituents out of equilibrium).  
Our models show that the reactions previously used to predict quenching of CO and CH$_{4}$ on giant planets and brown dwarfs may not be the rate-limiting steps in the quenching process.  Because of this result and previously described problems with mixing length scale assumptions \citep{smith1998}, the time-scale arguments require several updates, which are presented here.  Due to the overall uncertainties in the $\textrm{CO}\rightleftarrows \textrm{CH}_{4}$ 
kinetic pathways, the rate-limiting reactions identified by our thermochemical kinetics and 
transport model may change with advances in our understanding of C-H-O kinetics at high 
temperatures under reducing conditions.  As such, we also outline procedures that can be used 
to help derive $\tau_{chem}$ for any other potential rate-limiting reaction in the 
$\textrm{CO}\rightleftarrows \textrm{CH}_{4}$ interconversion mechanism, specifically for cases in which existing rate coefficients need to be reversed.  To illustrate the use of these time-scale arguments, transport-induced quenching of CO in the atmosphere of the T dwarf Gliese 229B and the quenching of CH$_4$ in the atmosphere of the hot-Jupiter HD 189733b become our specific examples to be discussed in detail.   

Our primary objectives are to 1) describe a method for correctly reversing 
chemical reactions using the principle of microscopic reversibility, 2) identify the dominant 
chemical pathways for the quenching of CO and CH$_4$ on Gliese 229B and HD 189733b, respectively, as determined 
from the results of the kinetics-and-transport model, and 3) describe revisions to simple time-scale
arguments that can be used to more accurately predict the quenched abundances of either CO or CH$_4$
in substellar objects.  We also revise our estimate of the water abundance in Jupiter's deep troposphere \citep{visscher2010icarus} based upon updates to the $\textrm{CO}\rightarrow \textrm{CH}_{4}$ kinetic scheme.  Throughout the following, we focus on thermochemistry and mixing only, without photochemistry, in order to specifically examine the  effect of atmospheric transport on the abundances of key atmospheric constituents.

\section{Chemical Model}\label{s:Chemical Model}

\subsection{Numerical Approach}\label{ss:Numerical Approach}

Our chemical model is based upon a thermochemical kinetics and diffusion model developed for 
Jupiter \citep{visscher2010icarus,moses2010} and for the hot-Jupiter exoplanets HD 189733b 
and HD 209458b \citep{moses2011}.  This one-dimensional (1-D) model, which uses the Caltech/JPL 
KINETICS code \citep{allen1981}, is described more completely in the above citations; here, 
we briefly discuss the salient points.  The code uses finite-difference techniques to solve 
the continuity equations controlling the vertical distribution of neutral carbon-, oxygen-, and
nitrogen-bearing constituents.  Because 1-D models cannot capture the complex three-dimensional 
transport processes that result from convection, atmospheric waves, and eddies of all sizes, 
all such vertical mixing processes are typically approximated in 1-D models as ``eddy diffusion'' parameterized by the eddy diffusion coefficient $K_{zz}$.  Free-convection and mixing-length theories 
\citep{stone1976,flasar1977} help constrain the magnitude of $K_{zz}$ in the deep atmospheres 
of giant planets and brown dwarfs; however, the constraints are loose, and we consider the 
$K_{zz}$ profile to be a free parameter in our models.  Using mixing-length theory and previous estimates of $K_{zz}$ \citep[e.g.,][]{griffith1998,griffith1999,cooper2006,showman2009,lewis2010,youdin2010} as a guide, we examine the sensitivity of our 
results to variations in $K_{zz}$ values ranging from $10^{3}$ to $10^{9}$ cm$^{2}$ s$^{-1}$ 
on Gliese 229B and from $10^{7}$ to $10^{11}$ cm$^{2}$ s$^{-1}$ on HD 189733b.  The Gliese 229B 
results described here are new, and a full description of the model results will be presented in a
later publication; further details of the HD 189733b model, along with comparisons to other exoplanet chemical models \citep[e.g.,][]{liang2003,liang2004,zahnle2009,zahnle2009arxiv,line2010} can be found in \citet{moses2011}.  

\begin{figure}
\begin{centering}
\scalebox{0.4}{\includegraphics[angle=0]{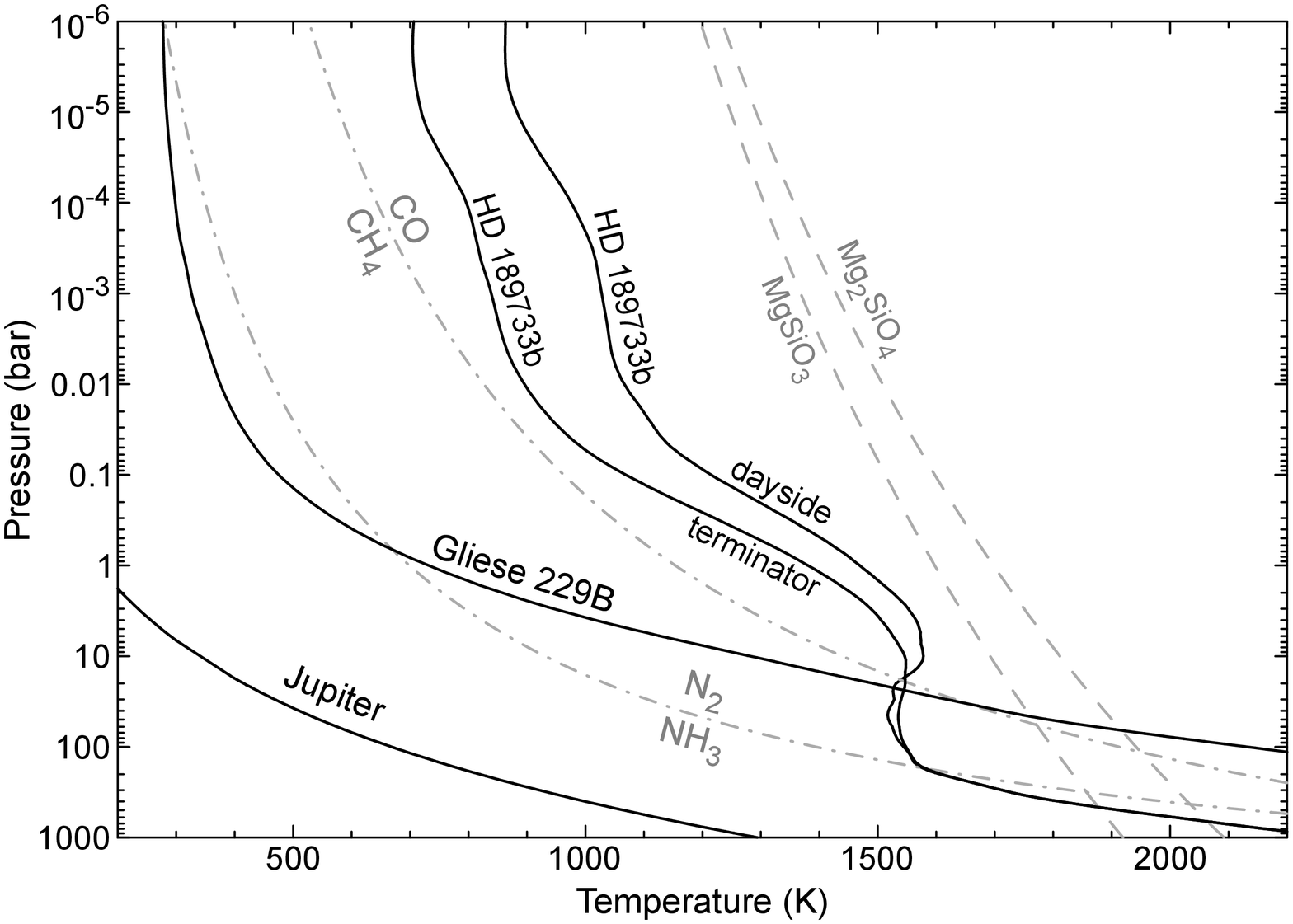}} \caption[Atmospheric
Profiles]{\footnotesize Pressure-temperature profiles (solid lines) used in our models for Gliese 229B, based upon Model B in \citet{saumon2000} and dayside-average and terminator-average profiles for HD 189733b from \citet{moses2011}, based upon
the GCM results of \citet{showman2009}.  The atmospheric profile for Jupiter is shown for comparison.  Also shown are
thermochemical equilibrium equal-abundance boundaries (dash-dotted lines) for major nitrogen (NH$_{3}$, N$_{2}$) and carbon (CH$_{4}$, CO) gases and the equilibrium condensation curves (dashed lines) for forsterite (Mg$_{2}$SiO$_{4}$) and enstatite (MgSiO$_{3}$) (dashed lines) in a solar-metallicity gas \citep{lodders2009}.}\label{figure:profiles}
\end{centering}
\end{figure}

The pressure-temperature profiles adopted for our models are shown in Figure 
\ref{figure:profiles}.  The profile for Gliese 229B is taken from Model B of \citet{saumon2000}.  
The profiles for HD 189733b are from \citet{moses2011} and represent dayside-average and terminator-average thermal structures based upon the 3-D GCM simulations of \citet{showman2009}.  These profiles are chosen for their relevance to atmospheric conditions at secondary eclipse and primary transit, respectively, and include extensions to 
low pressures using the thermospheric models of \citet{garciamunoz2007} and extensions to 
high pressures using the 1D models of \citet{fortney2006dyn}.  We adopt protosolar 
elemental abundances from \citet{lodders2009}, with an assumed 0.5$\times$ protosolar composition 
($[\textrm{Fe/H}]=-0.3$) for Gliese 229B based upon Model B in \citet{saumon2000}, and 
an assumed 1$\times$ protosolar composition ($[\textrm{Fe/H}]=0$) for HD 189733b.  For both objects, 
we assume uniform metallicity for all elements ({e.g.}, solar C/O ratio) and consider the removal 
of 20\% of oxygen by reaction with rock-forming elements in the deep atmosphere \citep[e.g., note silicate condensation curves in Fig.~\ref{figure:profiles}; see][]{lodders2004apj,visscher2005}.  Thermochemical equilibrium 
calculations are used to determine the initial atmospheric composition in the absence of transport, as 
is described in \citet{visscher2010icarus} and \citet{moses2010}.  

\subsection{Reaction Kinetics}\label{ss:Reverse Reactions}

The reaction list in our thermochemical kinetics and transport model is similar 
to that described in \citet{visscher2010icarus} and \citet{moses2010,moses2011}.
All of the chemical reactions in the model are reversed using the principle of microscopic 
reversibility.  As an aid to other investigators, we discuss this approach in detail to  
show how reverse rate coefficients are correctly calculated, particularly for reactions with an unequal
number of reactants and products.  For example, if we consider a 
balanced, elementary, gas-phase chemical reaction,
\begin{equation}\label{eq:gasreaction}
a\textrm{A}+b\textrm{B} \rightleftarrows c\textrm{C}+d\textrm{D},
\end{equation}
there are $a$ molecules of species A, $b$ molecules of species B, and so on, and the 
expression for the equilibrium constant $K_{eq}$ of the reaction can be written as
\begin{equation}\label{eq:eqcon}
K_{eq}=\frac{[\textrm{C}]^{c}[\textrm{D}]^{d}}{[\textrm{A}]^{a}[\textrm{B}]^{b}},
\end{equation}
where $[i]$ is the number density (molecules cm$^{-3}$) of each species $i$. The rates ($\nu$) of 
the forward reaction and reverse reaction are given by
\begin{align}
\nu_{f} &= k_{{f}}[\textrm{A}]^{a}[\textrm{B}]^{b}\\
\nu_{r} &= k_{{r}}[\textrm{C}]^{c}[\textrm{D}]^{d}
\end{align}
where $k_{f}$ and $k_{r}$ are the kinetic rate coefficients for the forward and reverse reactions, 
respectively.  When the system is operating at equilibrium, the forward reaction rate is equal 
to the reverse reaction rate:
\begin{equation}
k_{{f}}[\textrm{A}]^{a}[\textrm{B}]^{b}=k_{{r}}[\textrm{C}]^{c}[\textrm{D}]^{d}
\end{equation}
which is an expression of the principle of microscopic reversibility.  Substitution into the equilibrium constant expression (\ref{eq:eqcon}) gives an equation often found in textbooks,
\begin{equation}\label{eq:reversibility}
K_{eq}=\frac{k_{f}}{k_{r}},
\end{equation}
that can be used to determine reverse reaction rate coefficients.  However, it is important to note that thermodynamic tabulations containing equilibrium constants ($K$) are generally calculated assuming units of pressure \citep[e.g., 1-bar constant-pressure reference state,][]{chase1998}, whereas reaction rate coefficients ($k$) are generally reported in units of number density.  For reactions containing different numbers of reactants and products, omission of the appropriate pressure-correction terms in equation (\ref{eq:reversibility}) will result in incorrect rate coefficients for the reverse reaction.  

For example, several authors consider $\textrm{H}+\textrm{H}_{2}\textrm{CO}+\textrm{M}\rightarrow 
\textrm{CH}_{3}\textrm{O} + \textrm{M}$ as the rate-limiting step for $\textrm{CO}\rightarrow 
\textrm{CH}_{4}$ conversion in substellar atmospheres 
\citep{yung1988,griffith1999,bezard2002,cooper2006,line2010}, where M represents any third body.  
The rate coefficient for this termolecular reaction has not been measured experimentally, but it 
can be calculated from the rate coefficient of the reverse reaction $\textrm{CH}_{3}\textrm{O} + 
\textrm{M}$ $\rightarrow$ H + H$_2$CO + M investigated by \citet{page1989}. As pointed out by 
\citet{bezard2002}, the pressure-correction term was omitted by \citet{griffith1999} in their 
reversal, and therefore \citet{griffith1999} have adopted an incorrect rate coefficient in their 
treatment of CO quench kinetics on Gliese 229B.  \cite{line2010} make a similar error in their 
treatment of quench kinetics on HD 189733b and derive a two-body rate coefficient \citep[which was subsequently adopted by][for GJ 436b]{madhusudhan2011} for the 
three-body reaction 
$\textrm{H}+\textrm{H}_{2}\textrm{CO}+\textrm{M}$.  However, because 
\citet{line2010} consider a quench level that is near 1 bar, their omission of 
the pressure-correction term is mostly offset by their omission of the [M] term 
when calculating the rate of $\textrm{H}+\textrm{H}_{2}\textrm{CO}+\textrm{M}$.  
Nevertheless, such errors result in inaccurate reaction-rate estimates and may therefore result 
in incorrect conclusions regarding quench abundances and the strength atmospheric mixing on 
substellar objects \citep[e.g.,][]{griffith1999,line2010,madhusudhan2011}, particularly for cases 
in which the chemical behavior of an observed quenched species is used to constrain the value of 
the eddy diffusion coefficient.

In order to correctly reverse reactions with different numbers of products and reactants, we 
derive the appropriate pressure-correction term for equation (\ref{eq:reversibility}).  The 
equilibrium constant expression for reaction (\ref{eq:gasreaction}) in terms of pressure is 
written as
\begin{equation}\label{eq:Kp expression}
K_{P}=e^{-\Delta_{r} G^{\circ}/RT}=\frac{P_{\textrm{C}}^{c}P_{\textrm{D}}^{d}}{P_{\textrm{A}}^{a}P_{\textrm{B}}^{b}}
\end{equation}
where $R$ is the gas constant, $P_{i}$ is the partial pressure (in bars) of each species $i$, and $\Delta_{r} G^{\circ}$ is the Gibbs free energy change for the reaction (J mol$^{-1}$) at the standard-state pressure \citep[e.g., $P^{\circ}=1$ bar;][]{chase1998}, calculated from the Gibbs free energy of formation of the products and reactants:
\begin{equation}
\Delta_{r} G^{\circ} = \Delta_{f} G^{\circ} (\textrm{products}) - \Delta_{f} G^{\circ} (\textrm{reactants}).
\end{equation}
A discussion of the relation between $K_{p}$ and $\Delta_{r} G^{\circ}$ in equation (\ref{eq:Kp expression}) can be found in various thermodynamics texts \citep[e.g., Appendix I in][]{lewis2004} .  Note that $P_{i}$ is sometimes divided by the standard-state pressure of $P^{\circ}=1$ bar, so that $K_{P}$ remains a dimensionless quantity for reactions that have different numbers of products and reactants. The partial pressure of each 
constituent is a function of its mole fraction abundance $X_{i}$ and the total pressure $P_{T}$,
\begin{equation}\label{eq:number density}
P_{i}=X_{i}P_{T}=\frac{[i]}{n}P_{T}.
\end{equation}
The total number density of the system is calculated for a given volume from the ideal gas law,
\begin{equation}\label{eq:ideal gas law}
n=\frac{P_{T}VN_{A}}{RT}=\frac{P_{T}}{k_{B}T},
\end{equation}
in which $N_{A}$ is Avagadro's number and $k_{B}$ is Boltzmann's constant ($R/N_{A}$).  Substituting 
and rewriting equation (\ref{eq:Kp expression}) in units of number densities from equation 
(\ref{eq:number density}) gives
\begin{equation}
K_{P}=\frac{[\textrm{C}]^{c}[\textrm{D}]^{d}}{[\textrm{A}]^{a}[\textrm{B}]^{b}}\left(\frac{P_{T}}{n}\right)^{(c+d-a-b)}
= K_{eq} \left(\frac{P_{T}}{n}\right)^{(c+d-a-b)},
\end{equation}
where $P_{T}$ is units of bars and [$i$] and $n$ are in units of molecules cm$^{-3}$.  Rearrangement of 
this expression and substitution from equations (\ref{eq:reversibility}) and (\ref{eq:ideal gas 
law}) yields a general equation for calculating the reverse rate coefficient from the forward 
rate coefficient:
\begin{equation}\label{eq:reversal}
k_{r}=\frac{k_{f}}{e^{-\Delta_{r}G^{\circ}/RT}}({1.38065\times10^{-22}T})^{(n_{p}-n_{r})},
\end{equation}
for $T$ in kelvins, and where $n_{p}$ and $n_{r}$ are the numbers 
of products and reactants, respectively, in the
forward reaction.  Note that in cases where the number of products equals the number of reactants
({\em i.e.}, when $n_{p}-n_{r}=0$), the pressure-correction term becomes unity and
$K_{P}=K_{eq}=k_{f}/k_{r}$, as in equation (\ref{eq:reversibility}).  

For three-body (termolecular) reactions, the forward rate coefficient $k_{f}$ used in equation 
(\ref{eq:reversal}) can be obtained from experimental or theoretical data at appropriate
temperatures and pressures, and $k_r$ can then be determined from the above procedure at each
temperature-pressure point along the atmospheric grid.  If $k_f$ has not been measured at each
$P$-$T$ point along the grid, as is often the case, approximate expressions can be used.  For
example, the rate coefficient at the low-pressure limit ($k_0$ in units of cm$^{6}$ s$^{-1}$) 
and high-pressure limit ($k_{\infty}$ in units of cm$^{3}$ s$^{-1}$) have often been determined 
at a function of temperature.  At intermediate pressures, $k_{f}$ can be calculated from the expression
\begin{equation}\label{eq:threebody}
k_{f}=\frac{k_{0}}{1+({k_{0}[\textrm{M}]}/{k_{\infty}})}F_{c}^{\beta},
\end{equation}
where $\beta$ is given by
\begin{equation}
\beta  = \left(1+\left[\frac{\log_{10}(k_{0}[\textrm{M}]/k_{\infty})}{0.75-1.27\log_{10}F_{c}}\right]^{2}\right)^{-1},
\end{equation}
and $F_{c}$ is the center broadening factor \citep[e.g., see][]{baulch2005}.  

As in our previous studies \citep{visscher2010icarus,moses2010}, we calculate $k_f$ for every
pressure and temperature along the atmospheric grid, and then use equation (\ref{eq:reversal}) 
to determine $k_{r}$ for the reverse of every reaction at each atmospheric level using the 
appropriate temperature- and pressure-dependent values for $k_{f}$ and $\Delta_{r}G^{\circ}$.  The outcome  of this approach is a rate coefficient list of $\sim$800 forward reactions and $\sim$800 
corresponding reverse reactions involving H-C-N-O species, for each atmospheric profile.  As a 
result --- when not including photochemistry and atmospheric transport, and given enough time to 
achieve steady state --- our fully-reversed kinetics model yields results indistinguishable from those 
given by thermochemical-equilibrium calculations.  Further description of the model can be found in
\citet{visscher2010icarus} and \citet{moses2010,moses2011}.

\section{Model Results}
\subsection{CO Quench Chemistry on Gliese 229b}\label{ss: CO Quench Chemistry}
\begin{figure}
\begin{centering}
\scalebox{0.4}{\includegraphics[angle=0]{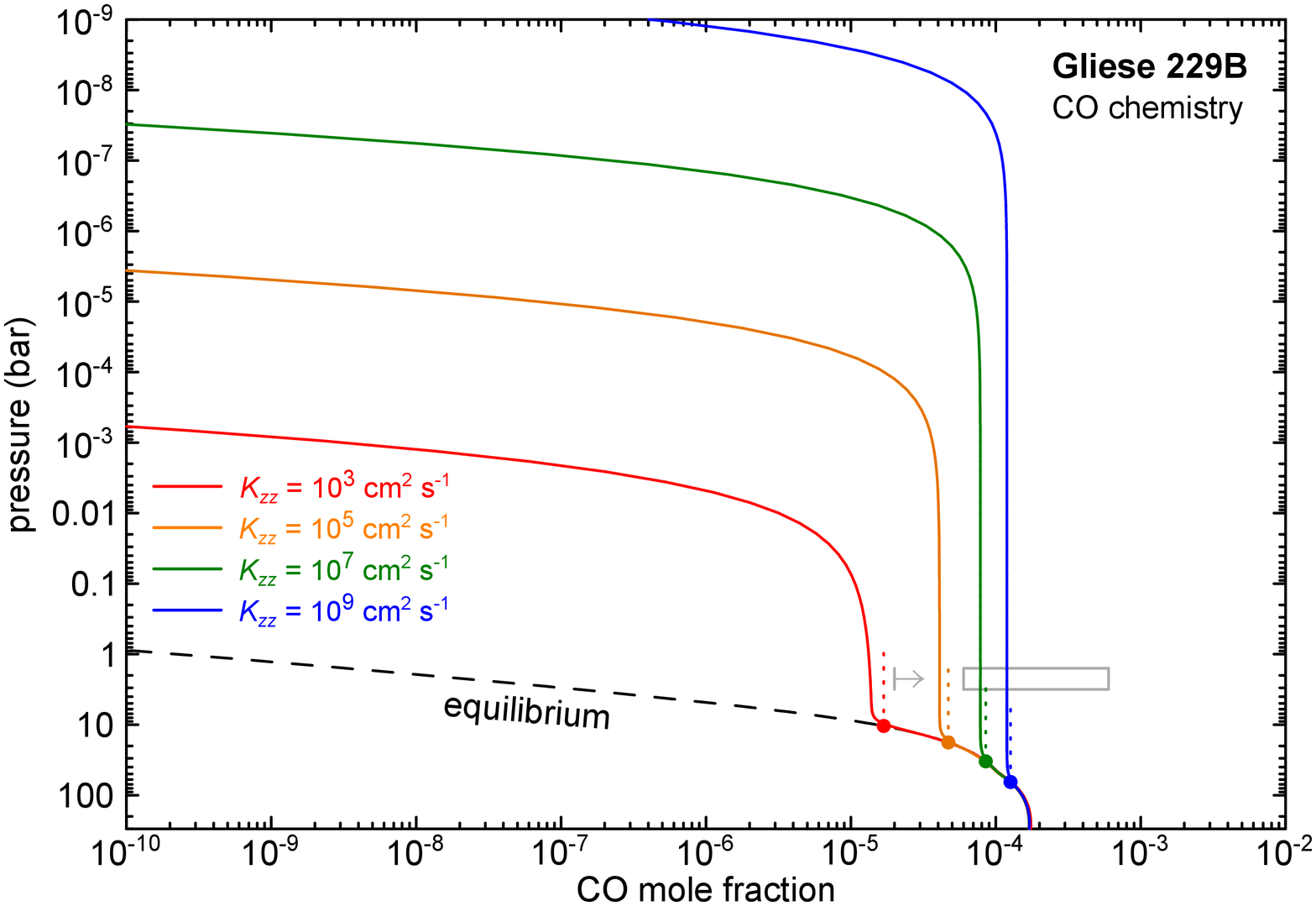}} \caption[Monoxide Quench Chemistry]{\footnotesize CO 
chemistry in the atmosphere of Gliese 229B for models considering thermochemical equilibrium 
only (dashed line) and thermochemistry with transport (solid lines) assuming $K_{zz}$ values of 
$10^{3}$, $10^{5}$, $10^{7}$, and $10^{9}$ cm$^{2}$ s$^{-1}$ and a 0.5$\times$ protosolar 
($[\textrm{Fe/H}]=-0.3$) composition \citep{saumon2000,lodders2009}.  For each case, the drop in 
the in the CO abundance at very low pressures illustrates where molecular diffusion begins to 
dominate over eddy diffusion.   The gray box shows a CO mole fraction of $60-600$ 
ppm based upon the 4.7 $\mu$m observations of \citet{noll1997} and \citet{oppenheimer1998} for metallicities ranging from $[\textrm{Fe/H}] = -0.5$ to $-0.1$ in the models of \citet{saumon2000}. The vertical gray bar with arrow 
represents a lower limit to the CO abundance of $X_{\textrm{CO}}\geq 20$ ppm from an analysis 
of the \citet{noll1997} data by \citet{griffith1999}, assuming a metallicity of 
$[\textrm{Fe/H}]\approx-0.6$.  The filled 
circles with dotted lines show the quench level (where $\tau_{chem}=\tau_{mix}$) and quenched CO mole 
fraction, respectively, as derived from the updated time-scale approach for each value of 
$K_{zz}$.}\label{figure:monoxide}
\end{centering}
\end{figure}

Results from our kinetics and transport model for Gliese 229B are given in Figure 
\ref{figure:monoxide}, which shows the CO abundance profiles for a $0.5\times$ 
protosolar-composition gas ($[\textrm{Fe/H]}\, =\, -0.3$) and various assumptions about the 
strength of vertical mixing (characterized by $K_{zz}$ values from $10^{3}$ to $10^{9}$ 
cm$^{2}$ s$^{-1}$).  The dashed line 
indicates the thermochemical equilibrium mole fraction abundance of CO calculated along the
atmospheric profile using the NASA CEA code \citep{gordon1994}.  Also shown are determinations 
of the CO mole fraction from ground-based 4.7 $\mu$m observations of Gliese 229B 
\citep[e.g.,][]{noll1997,oppenheimer1998,griffith1999,saumon2000}.  As predicted by 
\citet{fegley1996}, the observed CO abundance is several orders of magnitude larger than its 
equilibrium abundance in the observable regions of the atmosphere (see Figure \ref{figure:monoxide}), 
suggesting that CO is mixed upward from deeper regions where it is more favored thermodynamically.

In the deep atmosphere of Gliese 229B ($P_{T}\gtrsim 50$ bar), CO is the dominant carbon-bearing gas 
(cf.~Figure \ref{figure:profiles}), and its equilibrium abundance is maintained because the 
energy barriers for kinetic reactions are readily overcome at high temperatures.  At lower 
pressures, CH$_{4}$ replaces CO as the dominant carbon-bearing gas under equilibrium conditions, 
and the equilibrium CO mole fraction rapidly decreases with altitude.  However, at these colder 
altitudes, chemical reactions become more sluggish, and departures from equilibrium occur as 
atmospheric transport begins to dominate over reaction kinetics, eventually quenching the 
conversion of CO to CH$_{4}$.  As shown in Figure \ref{figure:monoxide}, the depth of the 
quench level (and consequently, the quenched CO abundance) depends upon strength of vertical 
mixing: for stronger mixing (characterized by larger $K_{zz}$ values) CO is quenched at 
relatively high pressures, deep in the atmosphere where it is more abundant; for weaker 
mixing (characterized by smaller $K_{zz}$ values), CO is quenched at lower pressures and higher 
altitudes where it is less abundant at equilibrium. Above the quench level, CO is mixed upward at 
a constant mole fraction until it starts rapidly decreasing where molecular diffusion 
begins to dominate over eddy diffusion.  In this scenario,
the atmosphere no longer remains well mixed, and heavy species like CO become confined to lower
altitudes.  For very low $K_{zz}$ values (e.g., $10^{3}$ cm$^{2}$ s$^{-1})$, the CO mole fraction 
does not remain constant above the quench level in our kinetics and transport model because 
molecular diffusion begins controlling atmospheric transport behavior at relatively high pressures.

As noted in \citet{visscher2010icarus} and \citet{moses2011}, we identify 
the main chemical pathways from the model results by comparing the relative rates of all reactions in the model.  Our results suggest that the dominant kinetic mechanism for $\textrm{CO}\rightarrow\textrm{CH}_{4}$ 
conversion in the troposphere of Gliese 229B consists of the following series of reactions:
\begin{subequations}\label{comech}
\begin{align}
\textrm{H} + \textrm{CO} & \xrightarrow{\textrm{M}} \textrm{HCO}\\
\textrm{H}_{2} + \textrm{HCO} & \rightarrow \textrm{H}_{2}\textrm{CO} + \textrm{H}\\[-1mm]
\textrm{H} + \textrm{H}_{2}\textrm{CO} & \xrightarrow{\textrm{M}} \textrm{CH}_{2}\textrm{OH}\\
\textrm{H}_{2} + \textrm{CH}_{2}\textrm{OH} & \rightarrow \textrm{CH}_{3}\textrm{OH}+\textrm{H}\\[-1mm]
\textrm{CH}_{3}\textrm{OH} & \xrightarrow{\textrm{M}} \textrm{CH}_{3} + \textrm{OH}\\
\textrm{H}_{2} + \textrm{CH}_{3} & \rightarrow \textrm{CH}_{4} + \textrm{H}\\
\textrm{H} + \textrm{OH} & \xrightarrow{\textrm{M}} \textrm{H}_{2}\textrm{O}\\[-3mm]
\cline{1-2}\textrm{CO} + 3\, \textrm{H}_{2} & \rightarrow \textrm{CH}_{4}+\textrm{H}_{2}\textrm{O},\tag{\ref{comech}, net} \end{align}
\end{subequations}
where M represents any third body.  

The slowest reaction in the fastest overall mechanism represents the rate-limiting step for CO $\rightarrow$ CH$_{4}$ conversion on Gliese 229B.  However, as discussed in \citet{visscher2010icarus}, there is a considerable degree of uncertainty in the identification of the rate-limiting step and the efficacy of the overall scheme, mostly due to uncertainties in laboratory and theoretical 
investigations of the product branching ratios for 
$\textrm{H}+\textrm{CH}_{3}\textrm{OH}\rightarrow \textrm{products}$ 
(see \citealt{visscher2010icarus} for details) and for $\textrm{CH}_{3}+\textrm{OH} \rightarrow \textrm{products}$ \citep[e.g., see][]{jasper2007}.  These uncertainties prompted \citet{moses2011} to use transition-state theory to calculate rate coefficients for the main $\textrm{H}+\textrm{CH}_{3}\textrm{OH}$ reaction pathways:
\begin{align}
\textrm{H}+\textrm{CH}_{3}\textrm{OH} \rightarrow \textrm{CH}_{2}\textrm{OH} + \textrm{H}_{2} & &k=1.09\times10^{-19}T^{2.728}e^{-2240/T}~\textrm{cm}^{3}~\textrm{s}^{-1}\\
\textrm{H}+\textrm{CH}_{3}\textrm{OH} \rightarrow \textrm{CH}_{3}\textrm{O} + \textrm{H}_{2} & &k=6.82\times10^{-20}T^{2.658}e^{-4643/T}~\textrm{cm}^{3}~\textrm{s}^{-1}\\
\textrm{H}+\textrm{CH}_{3}\textrm{OH} \rightarrow \textrm{CH}_{3} + \textrm{H}_{2}\textrm{O} & &k=4.91\times10^{-19}T^{2.485}e^{-10380/T}~\textrm{cm}^{3}~\textrm{s}^{-1}
\end{align}
Adopting updated reaction rate coefficients for $\textrm{H}+\textrm{CH}_{3}\textrm{OH}\rightarrow \textrm{products}$ from  \citet{moses2011} and for $\textrm{CH}_{3}+\textrm{OH} \rightarrow \textrm{products}$ from \citet{jasper2007}, we find that the thermal decomposition of methanol,
\begin{equation}
\textrm{CH}_{3}\textrm{OH} \xrightarrow{\textrm{M}} \textrm{CH}_{3} + \textrm{OH} \tag{\ref{comech}e}
\end{equation}
is the slowest reaction and is therefore the rate-limiting step for CO $\rightarrow$ CH$_{4}$ conversion in the above scheme (\ref{comech}).  This reaction mechanism differs from that originally proposed by \citet{prinn1977}, as well as that proposed by \citet{yung1988} and adopted by \citet{bezard2002} for Jupiter and by \citet{griffith1999} for Gliese 229B because the updated rate coefficients were not previously available.  The above scheme (\ref{comech}) also differs from that proposed by \citet{visscher2010icarus} for CO $\rightarrow$ CH$_{4}$ in Jupiter's troposphere, mostly due to differences in our selected reaction rate coefficients for $\textrm{H}+\textrm{CH}_{3}\textrm{OH} \rightarrow \textrm{CH}_{3} + \textrm{H}_{2}\textrm{O}$ \citep{moses2011} and $\textrm{CH}_{3} + \textrm{OH} \xrightarrow{\textrm{M}} \textrm{CH}_{3}\textrm{OH}$ \citep{jasper2007}.  Indeed, if we adopt the updated rate coefficients for our Jupiter models, the reaction scheme for CO destruction in Jupiter's troposphere is identical to the scheme (\ref{comech}) described above for Gliese 229B.  These revisions will also have some implications regarding the Jovian deep water abundance inferred from CO chemistry, which we briefly discuss in \S\ref{ss CO Jupiter} below.

In any case, the rate-limiting reaction candidates identified in \citet{visscher2010icarus} all 
quench in the same vicinity in the atmosphere of Gliese 229B and are therefore expected to 
yield roughly similar results for the quench CO abundance if our adopted rate coefficients 
for any of the reactions in scheme (\ref{comech}) are 
in serious error.  Furthermore, we emphasize that reaction (\ref{comech}e) is much more likely to be the 
rate-limiting step than the reaction $\textrm{H}_{2}+\textrm{H}_{2}\textrm{CO}\rightarrow 
\textrm{CH}_{3}+\textrm{OH}$ proposed by \citet{prinn1977}, which is too slow to 
play any significant role in CO quenching kinetics 
\citep[e.g.,][]{dean1987,yung1988,deavillezpereira1997,xia2001,krasnoperov2004,baulch2005,jasper2007} 
because there are faster, alternative pathways (such as scheme \ref{comech}) for CO destruction in 
hydrogen-dominated substellar atmospheres.

Although the overall reaction scheme and/or the rate-limiting reaction for 
$\textrm{CO}\rightarrow\textrm{CH}_{4}$ conversion may differ for objects with different 
compositions or thermal profiles, we expect the above scheme to play an important role in 
the atmospheres of relatively cool ({\em i.e.}, $X_{\textrm{CH}_{4}} \approx X_{\Sigma\textrm{C}}$) 
substellar objects (such as T dwarfs or cool giant planets) with near-solar metallicities and 
element abundance ratios (e.g., $\sim$ solar C/O).  However, under some conditions the CO quenching mechanism may bypass CH$_{3}$OH altogether via the following mechanism:
\begin{subequations}\label{cobypass}
\begin{align}
\textrm{H} + \textrm{CO} & \xrightarrow{\textrm{M}} \textrm{HCO}\\
\textrm{H}_{2} + \textrm{HCO} & \rightarrow \textrm{H}_{2}\textrm{CO} + \textrm{H}\\[-1mm]
\textrm{H} + \textrm{H}_{2}\textrm{CO} & \xrightarrow{\textrm{M}} \textrm{CH}_{2}\textrm{OH}\\
\textrm{H} + \textrm{CH}_{2}\textrm{OH} & \rightarrow \textrm{CH}_{3} + \textrm{OH}\\
\textrm{H}_{2} + \textrm{CH}_{3} & \rightarrow \textrm{CH}_{4} + \textrm{H}\\
\textrm{H} + \textrm{OH} & \xrightarrow{\textrm{M}} \textrm{H}_{2}\textrm{O}\\[-3mm]
\cline{1-2}\textrm{CO} + 3\, \textrm{H}_{2} & \rightarrow \textrm{CH}_{4}+\textrm{H}_{2}\textrm{O},\tag{\ref{cobypass}, net} \end{align}
\end{subequations}
in which the breaking of the C--O bond and the production of CH$_{3}$ + OH is again the rate-limiting step, in this case via the reaction
\begin{equation}
\textrm{H} + \textrm{CH}_{2}\textrm{OH}  \rightarrow \textrm{CH}_{3} + \textrm{OH}.\tag{\ref{cobypass}d}
\end{equation}
Note that the \emph{faster} of the two reactions (\ref{comech}e) or (\ref{cobypass}d) will serve as the rate-limiting step because this represents the fastest overall pathway available in our model for the CO $\rightarrow$ CH$_{4}$ conversion process.  However, although the thermal decomposition of methanol (reaction \ref{comech}e) remains the dominant rate-limiting reaction over the range of $K_{zz}$ values ($10^{3}-10^{9}$ cm$^{2}$ s$^{-1}$) shown in Figure \ref{figure:monoxide}, the contribution from reaction (\ref{cobypass}d) is significant enough (i.e., it is fast enough) that it must be considered when estimating the quench CO abundance via a timescale approach.

Using the kinetic schemes identified above, we revisit the time-scale approach that was previously developed
\citep[e.g.,][]{prinn1977,fegley1985apj,fegley1988,fegley1996,lodders2002} to estimate the quenched abundance of CO 
in the atmosphere of cool giant planets and brown dwarfs.  Considering reactions (\ref{comech}e) and (\ref{cobypass}d) as a combined rate-limiting step for CO destruction, the chemical lifetime of CO is given by
\begin{align}
\tau_{chem}\textrm{(CO)} & = \frac{[\textrm{CO}]}{-d[\textrm{CO}]/dt}\\
& = \frac{\textrm{[CO]}}{k_{\textrm{\ref{comech}e}}[\textrm{M}][\textrm{CH}_{3}\textrm{OH}]+k_{\ref{cobypass}d}[\textrm{H}][\textrm{CH}_{2}\textrm{OH}]}.
\end{align}
This expression is useful for considering the dominant contribution from either pathway breaking the C--O bond and forming CH$_{3}$ + OH.  For example, if reaction (\ref{comech}e) is much faster than reaction (\ref{cobypass}d), $\tau_{chem}(\textrm{CO})$ will be mostly determined by the rate of the methanol decomposition reaction.  The reaction rate coefficients $k_{\ref{comech}e}$ and $k_{\ref{cobypass}d}$ are calculated at each temperature in the model from their respective ``forward'' rate coefficients for $\textrm{CH}_{3}\textrm{OH} \xrightarrow{\textrm{M}} \textrm{CH}_{3} + \textrm{OH}$ and $\textrm{H} + \textrm{CH}_{2}\textrm{OH} \rightarrow \textrm{CH}_{3} + \textrm{OH}$ from \citet{jasper2007}, using equation (\ref{eq:reversal}) in the reversal procedure described above.  The parameters used for calculating the forward rate coefficients are given below 
in our discussion of CH$_{4}$ quench chemistry on HD 189733b.  

The vertical mixing time scale is given by
\begin{equation}\label{eq:tmix}
\tau_{mix}=\frac{L^{2}}{K_{zz}},
\end{equation}
where $K_{zz}$ is the eddy diffusion coefficient and $L$ is the characteristic length scale over 
which the mixing operates. Although the atmospheric pressure scale height $H$ has commonly been used 
for $L$ in time-scale comparisons, \citet{smith1998} has demonstrated that $L\approx H$ is not 
appropriate and that the mixing length $L$ is some fraction of $H$ that depends upon the thermal 
and eddy profiles of the substellar object, and upon the abundance profile of the atmospheric 
constituent (e.g., CO) under consideration.  Using the procedure outlined in \citet{smith1998}, 
\citet{bezard2002} and \citet{visscher2010icarus} have confirmed that $L\sim 0.1H$ for CO 
quenching kinetics in Jupiter's troposphere, and we find that $L\sim0.1H$ to $0.3H$ is appropriate 
for CO quenching kinetics on Gliese 229B.  A summary of results from our time-scale approach is 
given in Table \ref{tab: CO quench}, which lists $L/H$ ratios for CO quenching for different values of $K_{zz}$. 

The use of $L\approx H$ in chemical models involving 
$\textrm{CO}\rightleftarrows\textrm{CH}_{4}$ quenching kinetics 
(along with the use of incorrectly calculated reverse rate coefficients; see 
\S\ref{ss:Reverse Reactions}) may have serious implications for vertical mixing estimates on brown 
dwarfs such as Gliese 229B,
for which CO quench chemistry has been used to estimate the value of 
$K_{zz}$ \citep[][]{griffith1999}. For example, earlier investigators have constrained $K_{zz}$ on Gliese 229B to be in the range of $\sim10^{2}-10^{4}$
cm$^{2}$ s$^{-1}$ based upon previous suggestions of the rate-limiting step and assuming
$L\approx H$
\citep[e.g.,][]{griffith1999,saumon2000,leggett2007,mainzer2007}.  One explanation
for relatively low $K_{zz}$ values is that the CO quench level could be in the radiative zone where
convection no longer dominates, such that $K_{zz}$ could approach the sluggish values typically 
found in planetary lower stratospheres \citep[e.g.,][]{griffith1999,saumon2000,saumon2003iau}.  
However, the predicted negative temperature gradient \citep{saumon2000} at the CO quench point 
of a few tens of bars on Gliese 229B is larger than is typically found in stagnant, low-mixing 
regions in planetary atmospheres (in fact, temperature gradients are typically positive in such stagnant
regions where $K_{zz}$ is inferred to be in the $\sim$10$^{2}$--$10^{4}$ cm$^{2}$ s$^{-1}$ range 
in planetary atmospheres; see \citealt{yung1999}).  The low derived $K_{zz}$ values in the 
$\sim$10--100-bar region of Gliese 229B and other brown dwarfs are therefore surprising.  However, in 
contrast to previous investigations, our thermochemical kinetics and transport model and model-data 
comparisons shown in Figure \ref{figure:monoxide} suggest that the atmosphere of Gliese 229B is 
unlikely to be stagnant at the CO quench point.  We find that $K_{zz}$ 
values greater than 10$^7$ cm$^2$ s$^{-1}$ are needed to explain the observed CO mole fraction of 
60--600 ppm for assumed metalliticites that range from $[\textrm{Fe/H}]$ = -0.5 to -0.1 
(\citealt{saumon2000}, from an analysis of the 4.7 $\mu$m data of \citealt{noll1997} and 
\citealt{oppenheimer1998}), or $K_{zz}$ values greater 10$^4$ cm$^2$ s$^{-1}$ are needed for the 
observed lower limit on the CO mole fraction of 20 ppm, for an assumed metallicity of
$[\textrm{Fe/H}]$ = -0.6 \citep{griffith1999}.  In comparison, free-convection and mixing-length theories 
\citep{stone1976} predict a $K_{zz}$ values of 10$^8$ to 10$^9$ cm$^2$ s$^{-1}$ in the convection 
region of Gliese 229B.  Our results therefore do not preclude strong convective mixing at the 
CO quench point on brown dwarfs like Gliese 229B.

We emphasize that the inferences with regard to the CO abundance on brown dwarfs remain the 
same with our updated time-constant procedure.  It is simply the inferences with respect to 
$K_{zz}$ that have changed.  However, because assumptions about the strength of atmospheric
mixing can affect cloud models and other theoretical predictions of the vertical transport 
of condensates and gas-phase species \citep[e.g.,][]{golimowski2004,saumon2006,saumon2007,leggett2007,stephens2009,spiegel2009}, 
this change has important implications.  When our rate-limiting step for CO quenching is considered and the 
\citet{smith1998} effective length scale is adopted, the CO-based evidence for sluggish mixing 
in the $\sim$10--100-bar region of Gliese 229B (and potentially other brown dwarfs) 
disappears.

The quench level for CO is defined as the altitude for which $\tau_{chem}(\textrm{CO})=\tau_{mix}$.  In 
the time-scale approach, the quenched CO mole fraction that is mixed to higher altitudes 
is equal to the equilibrium abundance achieved at the quench level (characterized by a temperature 
$T_{q}$ and pressure $P_{q}$).  The results from our updated time-scale approach for CO quenching 
kinetics on Gliese 229B are also illustrated in Figure \ref{figure:monoxide}, where the filled 
circles with dotted lines indicate the quenched CO abundance for each value of $K_{zz}$.  In each 
case, the abundance estimated from the time-scale approach shows good agreement with the results 
from the full thermochemical kinetics and diffusion model (see Table \ref{tab: CO quench}).  
We therefore conclude that the time-scale approach provides a simple yet accurate method to 
describe the quench behavior of CO in the atmospheres of T dwarfs such as Gliese 229B --- provided that a reasonable rate limiting step and appropriate rate 
coefficient are used for calculating $\tau_{chem}$, and that the vertical mixing length scale $L$ 
advocated by \citet{smith1998} is used for calculating $\tau_{mix}$.  Note that because several plausible $\textrm{CO} \rightarrow \textrm{CH}_{4}$ rate-limiting reactions tend to quench in the same vicinity, quenching investigations using the approach of \citet{smith1998} to estimate $L$ give results similar to what would be expected from our kinetics and diffusion models \citep[e.g.,][]{bezard2002,cooper2006,saumon2006,saumon2007,geballe2009}.  For some brown dwarfs, these models \citep[e.g.,][]{saumon2006,saumon2007,geballe2009} indicate more sluggish mixing (i.e., lower $K_{zz}$ values) near the CO quench level than what our results suggest here for Gliese 229B.
 
\subsection{CH$_{4}$ Quench Chemistry on HD 189733b}\label{ss CH4 Quench Chemistry}

\begin{figure}
\begin{centering}
\scalebox{0.4}{\includegraphics[angle=0]{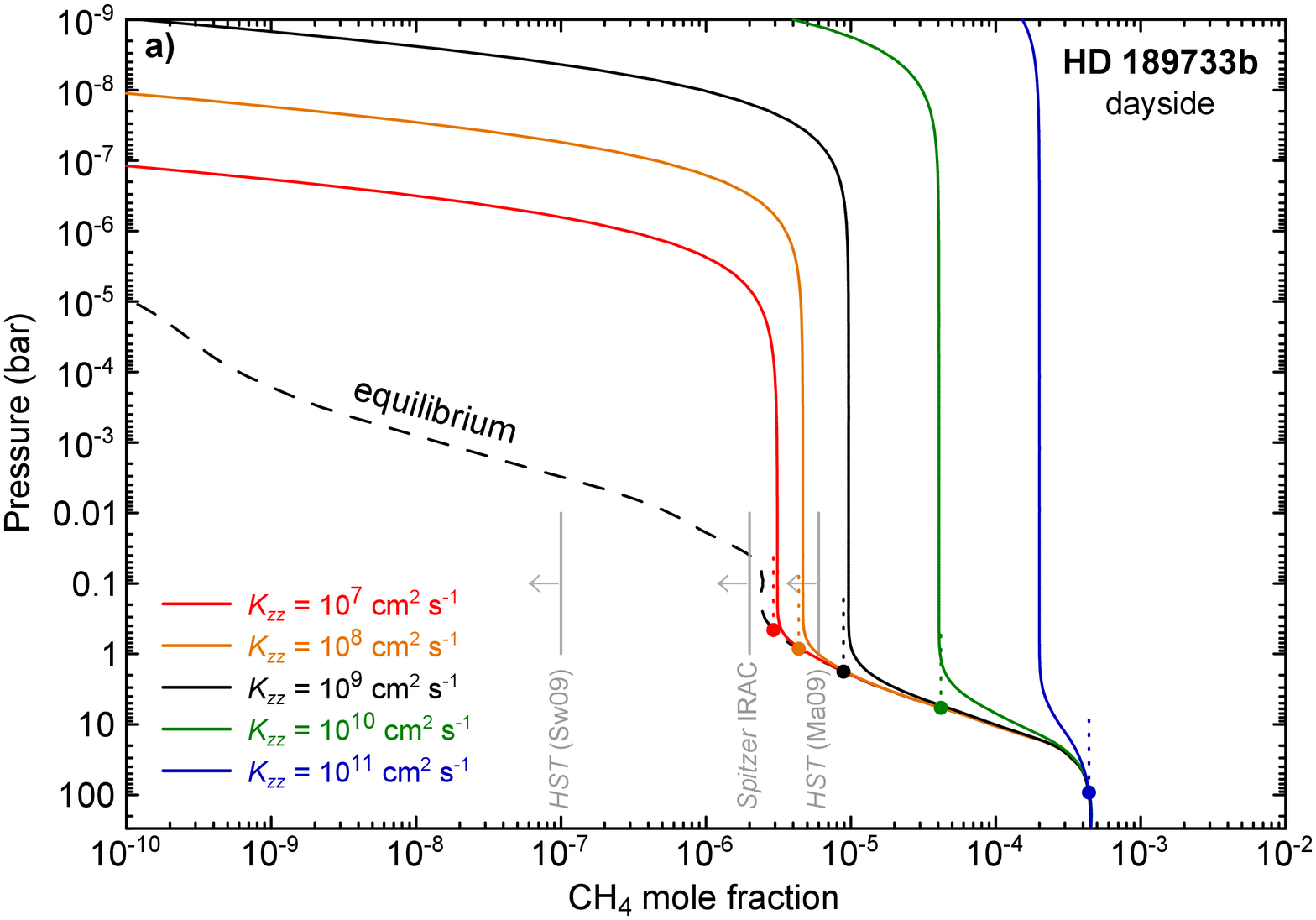}} \\
\scalebox{0.4}{\includegraphics[angle=0]{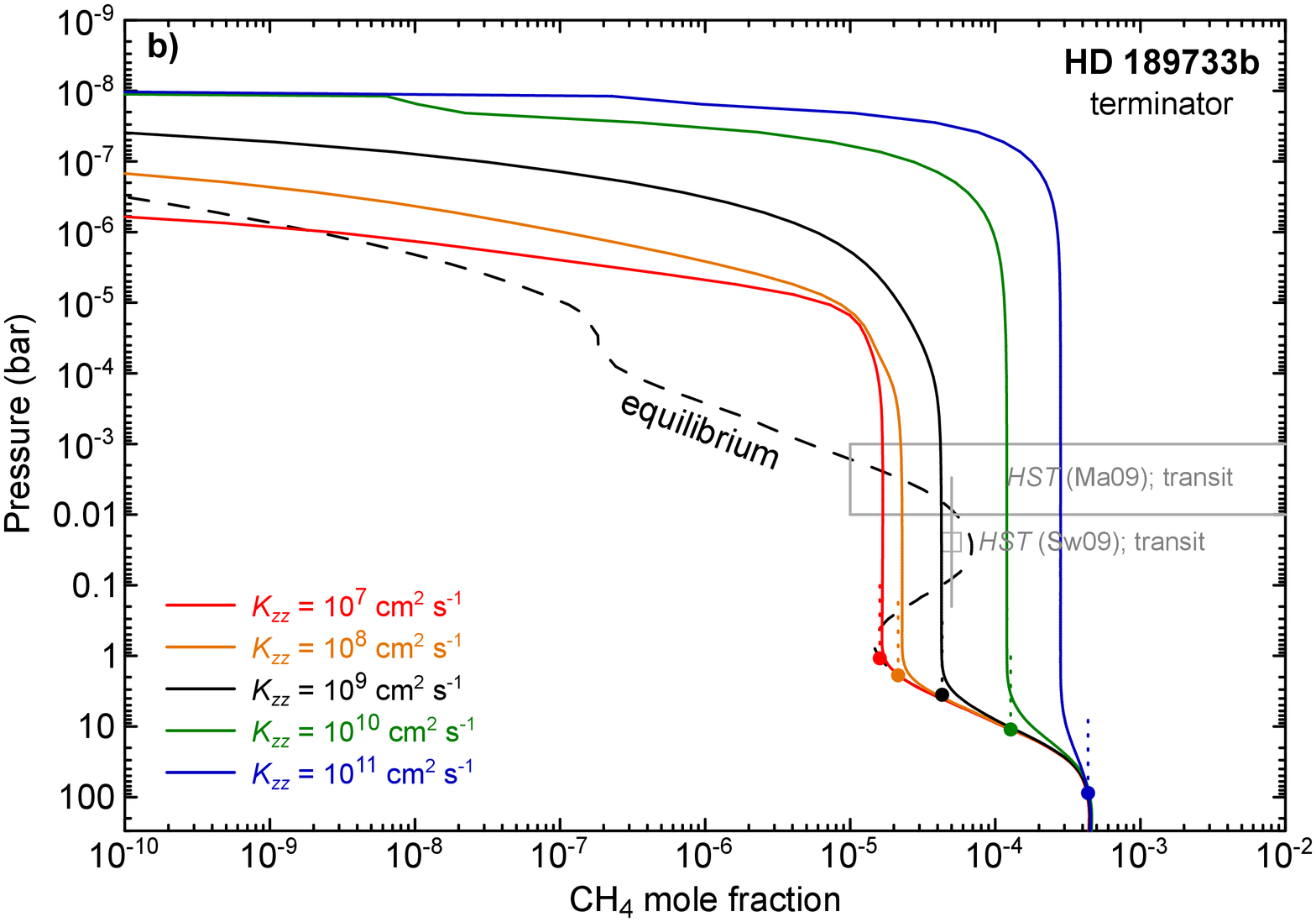}}
\small
\caption[Methane Quench Chemistry]{\footnotesize CH$_{4}$
chemistry in the atmosphere of HD 189733b \citep[profiles from][]{moses2011} for a) dayside-average models and b) terminator-average models, considering 
thermochemical equilibrium only (dashed line) and thermochemistry with transport (solid lines) 
assuming solar-metallicity composition \citep{lodders2009} and a range of $K_{zz}$ values (as labeled).  For each case, the drop-off in the CH$_{4}$ 
abundance at very low pressures illustrates where molecular diffusion begins to dominate over 
eddy diffusion.  In the dayside-average plot, the vertical gray bars with arrows represent upper limits on the methane abundance 
from \textit{HST\/}/NICMOS \citep{swain2009,madhusudhan2009} and \textit{Spitzer\/}/IRAC 
\citep{madhusudhan2009,charbonneau2008} observations for $0.01<P_{T}<1$ bar after \citet{swain2009}; 
upper limits from \citet{madhusudhan2009} are shown at similar pressures for comparison. In the terminator-average plot, detections 
of methane from \textit{HST\/}/NICMOS transit observations are indicated by gray boxes.  
\citet{swain2008} derive a best-fitting CH$_4$ mole fraction of $5\times10^{-5}$, whereas 
\citet{madhusudhan2009} find that CH$_{4}$/H$_{2}$ mixing ratios from $10^{-5}$ to 0.3 are 
consistent with the transmission spectra. The filled circles with dotted lines show the quench 
level (where $\tau_{chem}=\tau_{mix}$) and quenched CH$_{4}$ mole fraction, respectively, as derived from 
the time-scale approach for each value of $K_{zz}$.}\label{figure:methane}
\end{centering}
\end{figure}

The updated time-scale arguments also seem to be appropriate for describing the quenching 
of methane on HD 189733b.  Figure \ref{figure:methane} shows some results
from the thermochemical kinetics and diffusion modeling of \citet{moses2011} for an assumed 
solar-metallicity gas and variable assumptions about the rate of vertical transport (characterized 
by $K_{zz}$ values from $10^{7}$ to $10^{11}$ cm$^{2}$ s$^{-1}$) along the dayside-average and terminator-average profiles of HD 189733b.  Also shown are CH$_{4}$ upper limits from seconday eclipse observations (corresponding to our dayside-average models) made with the NICMOS instrument onboard the {\em Hubble\ Space\ Telescope\/}  
({\em HST}; \citealt{swain2008,swain2009, madhusudhan2009}) and with the IRAC instrument from the {\em Spitzer\
Space\ Telescope\/} \citep{charbonneau2008, madhusudhan2009}, as well as CH$_{4}$ abundance determinations during transit observations (corresponding to our terminator-average models) using {\em HST}/NICMOS \citep{swain2008,madhusudhan2009}.  

The dashed lines in Figure \ref{figure:methane} indicate the predicted thermochemical-equilibrium
abundance of CH$_{4}$, calculated using the NASA CEA code \citep{gordon1994}.  Methane is the 
dominant carbon-bearing gas in the deep atmosphere of HD 189733b ($P_{T} \gtrsim10$ bar), but 
its equilibrium abundance rapidly decreases with altitude as it is replaced by CO.  At high 
temperatures, the CH$_{4}$ abundance from the kinetics and transport models follows thermochemical 
equilibrium because forward and reverse reactions are relatively rapid (compared to mixing time 
scales) at high temperatures.  At lower temperatures, however, atmospheric transport begins to 
dominate over reaction kinetics, and the conversion of CH$_{4}$ to CO becomes quenched.
Disequilibrium abundances of CH$_{4}$ are therefore expected to be mixed into the upper 
atmosphere of HD 189733b \citep[see also][]{cooper2006,line2010,moses2011}.  As with CO 
quenching on Jupiter \citep{visscher2010icarus} and Gliese 229B (\S\ref{ss: CO Quench 
Chemistry}), the transition from equilibrium to a quenched regime is not abrupt but occurs 
over a range of altitudes roughly equal to one pressure scale height.  However, as we show 
for the $\textrm{CH}_{4} \rightleftarrows \textrm{CO}$ quenching cases, the assumption of a quench point where 
$\tau_{chem}=\tau_{mix}$ along the equilibrium profile (as in the time-scale approach) seems to
provide a reasonable approximation of the quenched abundance.

Comparison of the kinetics-with-transport model results for different $K_{zz}$ profiles in 
Figure \ref{figure:methane} illustrates the effect of the vertical mixing rate on CH$_{4}$ 
quench chemistry:  for stronger convective mixing (characterized by larger $K_{zz}$), CH$_{4}$ 
is quenched deeper in the atmosphere where it is more abundant; for weaker convective mixing 
(characterized by smaller $K_{zz}$), CH$_{4}$ is quenched at higher altitudes where it is 
less abundant.  For very large $K_{zz}$ values (e.g., $10^{11}$ cm$^{2}$ s$^{-1}$) 
methane will quench at deep altitudes where it is the dominant C-bearing gas and will 
therefore represent a major portion of the atmospheric carbon inventory.  In general, methane upper limits from the secondary eclipse observations are consistent with lower $K_{zz}$ values (i.e., $K_{zz} \lesssim 10^{8}$ cm$^{2}$ s$^{-1}$) whereas the transit detections of methane are consistent with higher values (i.e., $K_{zz} \gtrsim 10^{8}$ cm$^{2}$ s$^{-1}$).  Both photochemical destruction of CH$_{4}$ and transport-induced quenching on the warmer dayside would result in reduced CH$_{4}$ mole fractions on the dayside than at the terminators \citep{moses2011}.  However, the magnitude of these effects in the \citet{moses2011} models is inconsistent with the observed dayside-terminator differences, and one might also expect strong zonal winds to homogenize the CH$_{4}$ with longitude, leaving the observed differences unexplained.  As such, we cannot currently constrain $K_{zz}$ on HD 189733b.

As is discussed in more detail in \citet{moses2011}, CH$_{4}$ destruction in the deep atmosphere 
of HD 189733b occurs via the following series of reactions:
\begin{subequations}\label{ch4mech}
\begin{align}
\textrm{H}_{2}\textrm{O} & \xrightarrow{\textrm{M}} \textrm{OH} + \textrm{H}\\
\textrm{H} + \textrm{CH}_{4} & \rightarrow \textrm{CH}_{3} + \textrm{H}_{2}\\[-1mm]
\textrm{OH} + \textrm{CH}_{3} & \xrightarrow{\textrm{M}} \textrm{CH}_{3}\textrm{OH}\\
\textrm{H} + \textrm{CH}_{3}\textrm{OH} & \rightarrow \textrm{CH}_{2}\textrm{OH} + \textrm{H}_{2}\\[-1mm]
\textrm{CH}_{2}\textrm{OH} & \xrightarrow{\textrm{M}} \textrm{H}_{2}\textrm{CO} + \textrm{H}\\
\textrm{H} + \textrm{H}_{2}\textrm{CO} & \rightarrow \textrm{HCO} + \textrm{H}_{2}\\[-1mm]
\textrm{HCO} & \xrightarrow{\textrm{M}} \textrm{CO} + \textrm{H}\\[-3mm]
\cline{1-2} \textrm{CH}_{4}+\textrm{H}_{2}\textrm{O} & \rightarrow \textrm{CO}+3\textrm{H}_{2}, \tag{\ref{ch4mech}, net} \end{align}
\end{subequations}
where M refers to any third body.  This scheme is exactly the reverse of the 
$\textrm{CO}\rightarrow\textrm{CH}_{4}$ conversion scheme for CO quench kinetics in the atmosphere 
of Gliese 229B and Jupiter (see \S\ref{ss: CO Quench Chemistry}).  Using the revised rate coefficients for $\textrm{OH} + \textrm{CH}_{3}$ reaction pathways from \citet{jasper2007} and $\textrm{H} + \textrm{CH}_{3}\textrm{OH}$ reaction pathways from \citet{moses2011}, the rate-limiting step 
({i.e.}, the slowest reaction) for the above scheme on HD 189733b is the reaction
\begin{equation}
\textrm{OH} + \textrm{CH}_{3}  \xrightarrow{\textrm{M}} \textrm{CH}_{3}\textrm{OH}, \tag{\ref{ch4mech}c}
\end{equation}
which is the reverse of the rate-limiting step for CO destruction in the CH$_{4}$-dominated  atmosphere of Gliese 229B and Jupiter (\S\ref{ss: CO Quench Chemistry}).  The temperature-dependent rate-coefficient for reaction (\ref{ch4mech}c) is calculated via equation (\ref{eq:threebody}) using the modified Troe parameters given in \citet{jasper2007}:
\begin{align}
k_{0}  & = 1.932\times 10^{3} T^{-9.88}e^{-7544/T} + 5.109\times10^{-11}T^{-6.25}e^{-1433/T}~\textrm{cm}^{6}~\textrm{s}^{-1},\\
k_{\infty}  & = 1.031\times10^{-10}T^{-0.018}e^{16.74/T}~\textrm{cm}^{3}~\textrm{s}^{-1},\\
F_{c} & =0.1855 e^{-T/155.8}+0.8145 e^{-T/1675}+ e^{-4531/T},
\end{align}
As in the reverse scheme, under some conditions the CH$_{4}$ quenching process may bypass CH$_{3}$OH (i.e., reactions \ref{ch4mech}c and \ref{ch4mech}d) altogether, via the following mechanism :
\begin{subequations}\label{ch4bypass}
\begin{align}
\textrm{H}_{2}\textrm{O} & \xrightarrow{\textrm{M}} \textrm{OH} + \textrm{H}\\
\textrm{H} + \textrm{CH}_{4} & \rightarrow \textrm{CH}_{3} + \textrm{H}_{2}\\
\textrm{OH} + \textrm{CH}_{3} & \rightarrow \textrm{CH}_{2}\textrm{OH} + \textrm{H}\\[-1mm]
\textrm{CH}_{2}\textrm{OH} & \xrightarrow{\textrm{M}} \textrm{H}_{2}\textrm{CO} + \textrm{H}\\
\textrm{H} + \textrm{H}_{2}\textrm{CO} & \rightarrow \textrm{HCO} + \textrm{H}_{2}\\[-1mm]
\textrm{HCO} & \xrightarrow{\textrm{M}} \textrm{CO} + \textrm{H}\\[-3mm]
\cline{1-2} \textrm{CH}_{4}+\textrm{H}_{2}\textrm{O} & \rightarrow \textrm{CO}+3\textrm{H}_{2}, \tag{\ref{ch4bypass}, net} \end{align}
\end{subequations}
in which the slowest reaction is the formation of the C--O bond via the OH + CH$_{3}$ reaction,
\begin{equation}
\textrm{OH} + \textrm{CH}_{3}  \rightarrow \textrm{CH}_{2}\textrm{OH} + \textrm{H}.\tag{\ref{ch4bypass}c}
\end{equation}
The rate coefficient for reaction (\ref{ch4bypass}c) at each level in the model is calculated using equation (\ref{eq:threebody}) with modified Troe parameters from \citet{jasper2007}:
\begin{align}
k_{0}  & = 1.092\times 10^{-14} T^{0.996}e^{-1606/T} ~\textrm{cm}^{3}~\textrm{s}^{-1},\\
k_{\infty}  & = 5.864\times10^{-6}T^{5.009}e^{-949.4/T}~\textrm{s}^{-1},\\
F_{c} & =0.8622 e^{-T/9321}+0.1378 e^{-T/361.8}+ e^{-3125/T}.
\end{align}
As noted in \citet{moses2011}, the dominance of either reaction (\ref{ch4mech}c) or reaction (\ref{ch4bypass}c) as the rate-limiting step in the $\textrm{CH}_{4}\rightarrow \textrm{CO}$ mechanism depends upon the prevailing pressure and temperature conditions \citep[e.g.,][]{jasper2007}.  The CH$_{3}$OH-forming pathway (scheme \ref{ch4mech}) tends to dominate at higher pressures and higher $K_{zz}$ values ($\gtrsim 10^{9}$ cm$^{2}$ s$^{-1}$), whereas the alternative pathway (scheme \ref{ch4bypass}) tends to dominate at lower pressures and lower $K_{zz}$ values ($\lesssim 10^{9}$ cm$^{2}$ s$^{-1}$) on HD 189733b.  However, over the range of $K_{zz}$ values considered here, the relative rates of reactions (\ref{ch4mech}c) and (\ref{ch4bypass}c) are similar enough that both reactions should be considered when estimating the methane quench abundance in hot-Jupiter atmospheres.

\citet{line2010} were the first to examine CH$_{4}$ quench chemistry on HD 189733b by 
calculating $\tau_{chem}$(CO) for $\textrm{CO}\rightarrow \textrm{CH}_{4}$ conversion via the 
reaction $\textrm{H}+\textrm{H}_{2}\textrm{CO} \xrightarrow{\textrm{M}} \textrm{CH}_{3}\textrm{O}$ \citep{yung1988} and adopting $L=H$ for calculation of the vertical mixing time scale 
$\tau_{mix}$.  Aside from the omission of the pressure correction term when reversing 
$\textrm{CH}_{3}\textrm{O}+\textrm{M}$ (see \S\ref{ss:Reverse Reactions}), the use of $\tau_{chem}$(CO)
and the reaction rate for $\textrm{H}+\textrm{H}_{2}\textrm{CO}+\textrm{M}$ rather than its reverse 
is acceptable
 but not preferable for determining {\em methane\/} quench behavior via the time-scale approach 
(in which it assumed that 
$k_{f}[\textrm{H}][\textrm{H}_{2}\textrm{CO}][\textrm{M}]=k_{r}[\textrm{CH}_{3}\textrm{O}][\textrm{M}]$ 
until quenching occurs). In the context of applying the \citet{yung1988} kinetic scheme to HD 189733b, it would be more appropriate 
to calculate $\tau_{chem}$(CH$_{4}$) for the reverse reaction 
$\textrm{CH}_{3}\textrm{O} \xrightarrow{\textrm{M}} \textrm{H}_{2}\textrm{CO} + \textrm{H}$
because it is the characterization of CH$_4$ quenching that is the main objective.
Nevertheless, we recommend the use of the  $\textrm{OH} + \textrm{CH}_{3}$ reactions
$\textrm{OH} + \textrm{CH}_{3}  \xrightarrow{\textrm{M}} \textrm{CH}_{3}\textrm{OH}$ (\ref{ch4mech}c) and $\textrm{OH} + \textrm{CH}_{3} \rightarrow \textrm{CH}_{2}\textrm{OH} + \textrm{H}$ (\ref{ch4bypass}c) instead of $\textrm{CH}_{3}\textrm{O} \xrightarrow{\textrm{M}} \textrm{H}_{2}\textrm{CO} + \textrm{H}$ as the rate-limiting step for 
methane destruction in hot-Jupiter atmospheres, based upon a comparison of available $\textrm{CH}_{4}\rightarrow \textrm{CO}$ reaction pathways and recent updates to reaction kinetics \citep{moses2011}.

Using the reaction schemes (\ref{ch4mech}) and (\ref{ch4bypass}) described above, we can now test revisions to the time-scale approach for 
estimating the quenched CH$_{4}$ mole fraction in the atmosphere of HD 189733b.  Considering both $\textrm{OH} + \textrm{CH}_{3}$ pathways forming the C--O bond as a combined rate-limiting step provides a good estimate of the chemical lifetime for CH$_{4}$, given by
\begin{align}
\tau_{chem}(\textrm{CH}_{4}) & =\frac{[\textrm{CH}_{4}]}{-d[\textrm{CH}_{4}]/dt},\\
& = \frac{[\textrm{CH}_{4}]}{k_{\ref{ch4mech}c}[\textrm{M}][\textrm{CH}_{3}][\textrm{OH}]+k_{\ref{ch4bypass}c}[\textrm{CH}_{3}][\textrm{OH}]}.
\end{align}
The vertical mixing time scale ($\tau_{mix}$) is given by equation (\ref{eq:tmix}).
Using the procedure of \citet{smith1998}, we obtain $L\sim0.4H$ to $0.6H$ for CH$_{4}$ quenching 
kinetics in the atmosphere of HD189733b, depending upon the adopted $K_{zz}$ value \citep[cf.][for HD 209458b]{cooper2006}.  A summary of results from our time-scale approach for $K_{zz}=10^{7}-10^{10}$ cm$^{2}$ s$^{-1}$ is given in Table \ref{tab: CH4 quench}, which shows $L/H$  values for CH$_{4}$ quenching via for different $K_{zz}$ values in the  atmosphere of HD 189733b.  Note that the time-scale arguments generally compare very well (to within $\sim$10\% of) with the results of the thermochemical kinetics and transport model.  We therefore confirm the analytical approach of \citet{prinn1977} and conclude that the time-scale approach provides a simple yet accurate way to describe CH$_{4}$  quench chemistry in hot-Jupiter atmospheres --- again, provided that the chemical time scale $\tau_{chem}$ is calculated using the appropriate rate-limiting step ({\em i.e.}, reactions \ref{ch4mech}c and \ref{ch4bypass}c) and that the mixing time scale $\tau_{mix}$ is calculated using the appropriate vertical mixing length $L$ \citep{smith1998}.  However, the time-scale approximation begins to break down at very high $K_{zz}$ values (e.g., $10^{11}$ cm$^{2}$ s$^{-1}$; see Fig.~\ref{figure:methane}) in our HD 189733b models: although CO begins to depart from equilibrium once $\tau_{chem}>\tau_{dyn}$, the nearly isothermal behavior of the atmosphere in this region (cf.~Fig.~\ref{figure:profiles}) gives $\tau_{chem}\sim\tau_{dyn}$ over a wide range of altitudes above the quench level, so that there is no clear transition between the equilibrium and quench regimes.  In these cases, a kinetics and diffusion modeling approach is preferable for estimating the quench CH$_{4}$ abundance that is mixed into the upper atmosphere.

\subsection{Revised Estimate of Jupiter's Deep Water Abundance}\label{ss CO Jupiter}
As noted above, our preferred scheme for CO $\rightarrow$ CH$_{4}$ in Jupiter's troposphere differs from that proposed in previous studies of $\textrm{CO} \rightleftarrows \textrm{CH}_{4}$ quench kinetics \citep[e.g.][]{prinn1977,yung1988,bezard2002,cooper2006,visscher2010icarus,line2010,madhusudhan2011} because of updates to reaction kinetics. The rate limiting step for CO destruction is $\textrm{H}_{2} + \textrm{CH}_{3}\textrm{O} \rightarrow \textrm{CH}_{3}\textrm{OH} + \textrm{H}$ in the nominal model of \citet{visscher2010icarus}. Our current scheme (\ref{comech}) includes CH$_{2}$OH as an intermediate instead of CH$_{3}$O, and the formation of CH$_{3}$ (which rapidly reacts to form CH$_{4}$) via $\textrm{CH}_{3}\textrm{OH}  \xrightarrow{\textrm{M}} \textrm{CH}_{3} + \textrm{OH}$ (\ref{comech}e) instead of $\textrm{H} + \textrm{CH}_{3}\textrm{OH} \rightarrow \textrm{CH}_{3} + \textrm{H}_{2}\textrm{O}$.  As noted above, these differences from \citet{visscher2010icarus} are mostly due to revisions in our adopted reaction rate coefficients for $\textrm{H}+\textrm{CH}_{3}\textrm{OH} \rightarrow \textrm{CH}_{3} + \textrm{H}_{2}\textrm{O}$ \citep{moses2011} and $\textrm{CH}_{3} + \textrm{OH} \xrightarrow{\textrm{M}} \textrm{CH}_{3}\textrm{OH}$ \citep{jasper2007}, yielding a slower overall mechanism for CO $\rightarrow$ CH$_{4}$ on Jupiter.  Because CO quenching occurs at significantly higher pressures on Jupiter than on Gliese 229B, the reaction $\textrm{CH}_{3}\textrm{OH} \xrightarrow{\textrm{M}} \textrm{CH}_{3} + \textrm{OH}$ (\ref{comech}e) is the dominant rate-limiting step in Jupiter's troposphere.  For example, assuming $K_{zz}=10^{8}$ cm$^{2}$ s$^{-1}$, reaction (\ref{comech}e) is nearly two orders of magnitude faster than reaction (\ref{cobypass}d) at the quench level.  Estimates of the quench CO abundance on Jupiter via the timescale approach therefore need only to consider the methanol decomposition reaction (\ref{comech}e) as the rate-limiting step for calculating the CO chemical lifetime.

The mechanism and rate of CO destruction in Jupiter's troposphere has implications for the water abundance in Jupiter's interior, because the CO abundance is closely tied to the H$_{2}$O abundance via the net thermochemical reaction \citep[e.g.,][]{fegley1988}:
\begin{equation}
\textrm{H}_{2}\textrm{O} + \textrm{CH}_{4} = \textrm{CO} + 3\textrm{H}_{2}.
\end{equation}
Because our revised CO $\rightarrow$ CH$_{4}$ scheme is slower than that adopted by \citet{visscher2010icarus}, quenching occurs deeper in the troposphere where CO is more thermodynamically favored.   We therefore derive a lower estimate of the H$_{2}$O abundance than in \citet{visscher2010icarus}.  Adopting a CO mole fraction of $1.0\pm0.2$ ppb \citep{bezard2002} as an observational constraint for the internal/tropospheric CO source and considering a range of $K_{zz}$ values from $1\times10^{7}$ to $1\times10^{9}$ cm$^{2}$ s$^{-1}$, our updated model results yield a Jovian water abundance of 0.1-1.5 times the solar H$_{2}$O/H$_{2}$ ratio ($9.61\times10^{-4}$).  For comparison, the preferred model in \citet{visscher2010icarus} gives a water abundance of 0.4-3.4 times the solar H$_{2}$O/H$_{2}$ ratio, not including uncertainties in reaction kinetics.  
We derive the following empirical expression from the results of our kinetics and diffusion models:
\begin{equation}
X_{\textrm{CO}}=6.52\times10^{-13}K_{zz}^{0.443}E_{\textrm{H}_{2}\textrm{O}},
\end{equation}
which describes the relationship between the CO mole fraction ($X_{\textrm{CO}}$), the eddy diffusion coefficient ($K_{zz}$), and the water enrichment over the solar abundance ($E_{\textrm{H}_{2}\textrm{O}}$) in Jupiter's atmosphere, when $\textrm{CH}_{3}\textrm{OH} \xrightarrow{\textrm{M}} \textrm{CH}_{3} + \textrm{OH}$ (\ref{comech}e) is the rate-limiting step for CO destruction.

Our model results suggest that the subsolar water abundance ($0.51\times$ solar) measured by the \emph{Galileo} entry probe \citep{wong2004} is consistent with the observed CO abundance \citep{bezard2002}, and preclude formation mechanisms that would result in large water abundances in Jupiter's atmosphere (e.g., see \citealt{lodders2004apj} and \citealt{visscher2010icarus} for discussion).  If we take the \emph{Galileo} entry probe H$_{2}$O abundance ($0.51\times$ solar) as a lower limit and consider a factor-of-5 uncertainty in reaction kinetics for the rate-limiting step \citep[e.g.,][]{baulch2005}, our models are consistent with a Jovian water abundance of 0.51-2.6 times the solar abundance, corresponding to H$_{2}$O/H$_{2}\approx(4.9-25)\times10^{-4}$ in the deep atmosphere.  As noted above and in \citet{visscher2010icarus}, we emphasize that this estimate is subject to revision based upon advances in our understanding of $\textrm{CO} \rightarrow \textrm{CH}_{4}$ reaction kinetics, as well as improvements in the determination of $K_{zz}$ in Jupiter's troposphere.  Nevertheless, the plausible rate-limiting reactions for CO destruction generally imply lower water abundances than have been predicted from giant planet formation scenarios \citep[e.g.,][]{owen1999,owen2006,gautier2001,hersant2004,alibert2005,mousis2009}.

\section{Conclusions} \label{ss: Conclusions}

Our thermochemical kinetics and diffusion models for Gliese 229B and HD 189733b \citep[see also][]{moses2011} 
confirm that atmospheric transport strongly influences the chemical behavior of carbon-bearing species in 
the upper atmospheres of substellar objects.  Carbon monoxide is subject to transport-induced 
quenching on cool giant planets such as Jupiter and T dwarfs such as Gliese 229B, whereas methane 
may quench in the atmospheres of hot Jupiters such as HD 189733b.  From a comparison of the 
relative rates of all available chemical pathways for $\textrm{CO}\rightleftarrows \textrm{CH}_{4}$ 
interconversion in the thermochemical kinetics and transport
models presented here and elsewhere \citep{visscher2010icarus,moses2011}, we find that the destruction of the C--O bond
is the rate-limiting step for CO quenching in the CH$_{4}$-dominated objects, whereas formation of the C--O bond
is the rate-limiting step for CH$_{4}$ quenching in the CO-dominated objects.  Using updated reaction kinetics \citep{jasper2007,moses2011} the dominant forward/reverse reaction pair for the formation/destruction of the C--O bond is either $\textrm{OH} + \textrm{CH}_{3} + \textrm{M} \rightleftarrows \textrm{CH}_{3}\textrm{OH} + \textrm{M}$ or $\textrm{OH} + \textrm{CH}_{3} \rightleftarrows \textrm{CH}_{2}\textrm{OH} + \textrm{H}$, depending upon local temperature and pressure conditions near the quench level.  However, because of their similar rates, both reaction pathways should be considered together when calculating $\tau_{chem}$ for $\textrm{CO}\rightleftarrows \textrm{CH}_{4}$ quenching in brown dwarf or hot-Jupiter atmospheres.  This reaction mechanism differs from the forward/reverse reaction pair $\textrm{H}_{2} + \textrm{CH}_{3}\textrm{O} \rightleftarrows \textrm{CH}_{3}\textrm{OH} + \textrm{H}$ proposed by \citet{visscher2010icarus}, the forward/reverse reaction pair $\textrm{H}+\textrm{H}_{2}\textrm{CO}+\textrm{M}\rightleftarrows 
\textrm{CH}_{3}\textrm{O}+\textrm{M}$ adopted by \citet{yung1988} and subsequent authors 
\citep[e.g.,][]{griffith1999,bezard2002,cooper2006,saumon2006,hubeny2007,line2010,madhusudhan2011}, or the 
original suggestion of  H$_2$ + H$_2$CO $\rightleftarrows$ CH$_3$ + OH by 
\citet{prinn1977} and subsequent authors
\citep[e.g.,][]{fegley1985apj,fegley1988,fegley1996,lodders1994,lodders2002,saumon2003iau,visscher2005,hubeny2007} because of recent updates in reaction rate coefficients.  These revisions also have implications for estimates of the water abundance in Jupiter's deep troposphere using the CO observational constraint.  Using updated kinetics, our model results along with a lower limit provided by \emph{Galileo} entry probe measurements suggest a water abundance of approximately 0.51-2.6 times the solar abundance, corresponding to H$_{2}$O/H$_{2}\approx(4.9-25)\times10^{-4}$ in Jupiter's deep atmosphere.  The transport-induced quenching behavior of CO therefore implies lower H$_{2}$O abundances than have been predicted from several giant planet formation scenarios \citep[e.g., see][]{lodders2004apj,visscher2010icarus}.

For each substellar object, the rate of vertical transport (characterized by eddy diffusion coefficient $K_{zz}$) will strongly affect the 
quenched abundance of a given species: for higher $K_{zz}$ values, quenching occurs at deeper, 
hotter altitudes, whereas for lower $K_{zz}$ values, quenching occurs at higher, cooler altitudes.  Moreover, the equilibrium abundance of any atmospheric constituent at its quench level depends upon the bulk composition and thermal profile of the atmosphere ({\em i.e.}, the prevailing conditions at the quench point).  The detection and characterization of quench species may therefore provide constraints on the chemistry, structure, and mixing rates of substellar atmospheres \citep[e.g.,][]{fegley1996,lodders2002}.  For Gliese 229B, our results suggest significantly higher $K_{zz}$ rates near the CO quench level than have been previously inferred.  For HD 189733b, the terminator CH$_{4}$ detection by \citet{swain2008} and the corresponding analysis by \citet{madhusudhan2009} suggest high values of $K_{zz}$ ($\gtrsim 10^{8}$ cm$^{2}$ s$^{-1}$) at the $\textrm{CO} \rightleftarrows \textrm{CH}_{4}$ quench point, whereas CH$_{4}$ upper limits for the dayside atmosphere as observed during secondary eclipse \citep{charbonneau2008,madhusudhan2009,swain2008,swain2009} suggest lower values of $K_{zz}$ ($\lesssim 10^{8}$ cm$^{2}$ s$^{-1}$).  These differences imply that disequilibrium chemistry, local $K_{zz}$ differences, or other processes are complicating the simple picture of uniform transport-induced quenching in highly irradiated hot-Jupiter atmospheres.

We have presented an update to the time-constant procedure developed by \citet{prinn1977} and confirm  that their general analytical approach can be used to accurately describe the chemical 
behavior of quenched species.  In other words, full thermochemical kinetics and transport models are not needed to predict the quenched abundances of disequilibrium species, particularly for the case
of CO and CH$_4$ quenching, for which there is one specific quench point \citep[see
also][]{visscher2010icarus,moses2011}.  We find that the time-scale approach can be used for a 
wide range of substellar objects to estimate the expected mole fractions of any atmospheric 
constituents subject to transport-induced quenching, provided that 1) an appropriate 
dominant mechanism for chemical interconversion between constituents is selected, which may 
require a comparison of the relative rates of all chemical pathways contributing to the production 
and destruction of the species under consideration, 2) an appropriate rate-limiting reaction within 
that dominant mechanism is identified, and the rate coefficient for that reaction is determined
through laboratory or theoretical investigations (or proper reversal, if required) in order to 
calculate $\tau_{chem}$, and 3) an appropriate mixing length $L$ is adopted, following the procedure 
of \citet{smith1998}, for the calculation of $\tau_{mix}$. Previous questionable assumptions or 
incorrect applications of the time-constant procedure may have led to underestimates of the 
strength of atmospheric mixing at the CO quench level on Gliese 229B
and inaccurate determinations of 
the methane abundance on extrasolar giant planets.  Note that although the
time-constant approach is not limited to $\textrm{CO}\rightleftarrows \textrm{CH}_{4}$ but can
be applied to any chemical constituent subject to reaction kinetics and vertical mixing in 
substellar atmospheres (e.g., see the discussion of the quenching of nitrogen species on Jupiter by
\citealt{moses2010}), some constituents such as HCN or NH$_3$ may have complicated quench kinetics \citep{moses2011}, making the time-constant arguments difficult to apply in practice.  The time-scale approximation may also break down if $\tau_{chem}\sim\tau_{mix}$ over an extended pressure range, due to small temperature gradients (such as can be found in radiative regions) or other conditions.  As such, thermochemical kinetics and transport models may still have an important role to play in the prediction of transport-quenched abundances in the atmospheres of substellar objects.

\section*{Acknowledgements}
We thank Justin Troyer for contributing to preliminary kinetics and diffusion models of Gliese 229B.  This work was supported by the NASA Planetary Atmospheres Program (NNH08ZDA001N).  Support for C.V.~was also provided by the Lunar and Planetary Institute, USRA (NASA Cooperative Agreement NCC5-679).  LPI Contribution No.~1624.

\clearpage

\begin{deluxetable}{cccccccc} 
\tablecolumns{8} 
\tablewidth{0pc} 
\tablecaption{CO quench chemistry on Gliese 229B} 
\tablehead{
\colhead{} & \multicolumn{5}{c}{time-scale approach} & \colhead{} & \colhead{kinetics model}\\[2mm]
\cline{2-6} \cline{8-8} \\[-2mm]
\colhead{$K_{zz}$, cm$^{2}$ s$^{-1}$} & \colhead{$T_{q}$, K}   & \colhead{$P_{q}$, bar}    & \colhead{$\tau_{chem}(\textrm{CO})$, s} & \colhead{$L/H$} & \colhead{quenched $X_{\textrm{CO}}$} & \colhead{} & \colhead{quenched $X_{\textrm{CO}}$}}
\startdata 
$1\times10^{3}$ & 1295 & 10.4 & $4.16\times10^{6}$ & 0.140 & $1.69\times10^{-5}$ & & $1.33\times10^{-5}$\\
$1\times10^{5}$ & 1457 & 17.8 & $8.71\times10^{4}$& 0.181 & $4.70\times10^{-5}$& & $4.09\times10^{-5}$\\
$1\times10^{7}$ & 1651 & 33.2 & $1.82\times10^{3}$ & 0.230 & $8.53\times10^{-5}$& & $7.83\times10^{-5}$\\
$1\times10^{9}$ & 1910 & 64.9 & $3.29\times10^{1}$ & 0.268 & $1.26\times10^{-4}$& & $1.19\times10^{-4}$\\
\enddata 
\label{tab: CO quench}
\tablecomments{For the kinetics models, the CO mole fraction at the 1-bar level is adopted as the 
quenched abundance.}
\end{deluxetable} 

\begin{deluxetable}{cccccccc} 
\tablecolumns{8} 
\tablewidth{0pc} 
\tablecaption{CH$_{4}$ quench chemistry on HD 189733b} 
\tablehead{ 
\colhead{} & \multicolumn{5}{c}{time-scale approach} & \colhead{} & \colhead{kinetics model}\\[2mm]
\cline{2-6} \cline{8-8} \\[-2mm]
\colhead{$K_{zz}$, cm$^{2}$ s$^{-1}$} & \colhead{$T_{q}$, K}   & \colhead{$P_{q}$, bar}    &\colhead{$\tau_{chem}(\textrm{CH}_{4})$, s} & \colhead{$L/H$} & \colhead{quenched $X_{\textrm{CH}_{4}}$} & \colhead{} & \colhead{quenched $X_{\textrm{CH}_{4}}$}}
\startdata 
\sidehead{dayside-average thermal-structure models}
$1\times10^{7}$ & 1394 & 0.5 &  $7.86\times10^{6}$ & 0.383 & $2.92\times10^{-6}$ & & $3.11\times10^{-6}$\\
$1\times10^{8}$ & 1455 & 0.8 &  $1.09\times10^{6}$ & 0.435 & $4.37\times10^{-6}$ & & $4.66\times10^{-6}$\\
$1\times10^{9}$ & 1516 & 1.8 & $1.67\times10^{5}$ & 0.518 & $8.89\times10^{-6}$ & & $9.64\times10^{-6}$\\
$1\times10^{10}$ & 1573 & 5.8 & $2.62\times10^{4}$ & 0.630 & $4.18\times10^{-5}$ & & $4.05\times10^{-5}$\\
\sidehead{terminator-average thermal-structure models} 
$1\times10^{7}$ & 1389 & 1.1 &  $6.70\times10^{6}$ & 0.359 & $1.60\times10^{-5}$ & & $1.68\times10^{-5}$\\
$1\times10^{8}$ & 1448 & 1.9 &  $9.79\times10^{5}$ & 0.415 & $2.14\times10^{-5}$ & & $2.27\times10^{-5}$\\
$1\times10^{9}$ & 1506 & 3.6 & $1.53\times10^{5}$ & 0.502 & $4.31\times10^{-5}$ & & $4.25\times10^{-5}$\\
$1\times10^{10}$ & 1548 & 11.0 & $3.12\times10^{4}$ & 0.702 & $1.28\times10^{-4}$ & & $1.20\times10^{-4}$\\
\enddata 
\label{tab: CH4 quench}
\tablecomments{For the kinetics models, the CH$_{4}$ mole fraction at the 0.01-bar level is adopted as the quenched abundance.}
\end{deluxetable}

\clearpage

\clearpage

\clearpage

\end{document}